\documentclass[10pt,a4paper,twocolumn]{article}

\usepackage[T1]{fontenc}
\usepackage[utf8]{inputenc}
\usepackage{lmodern}

\usepackage[top=25mm,bottom=25mm,left=18mm,right=18mm]{geometry}
\usepackage{setspace}
\usepackage{balance}

\usepackage{amsmath,amssymb,amsfonts}
\usepackage{bm}
\usepackage{siunitx}

\usepackage{graphicx}
\usepackage{booktabs}
\usepackage{multirow}
\usepackage{array}
\usepackage{caption}
\usepackage{subcaption}
\usepackage{float}
\usepackage{tikz}
\usetikzlibrary{shapes.geometric,arrows.meta,positioning,fit,backgrounds}

\usepackage[version=4]{mhchem}   
\usepackage{chemformula}         
\usepackage{physics}             
\usepackage{gensymb}             

\usepackage[numbers,sort&compress]{natbib}
\usepackage{doi}
\usepackage{hyperref}
\hypersetup{
    colorlinks=true,
    linkcolor=blue,
    citecolor=blue,
    urlcolor=blue
}

\usepackage{titlesec}
\usepackage{authblk}
\usepackage{abstract}

\titleformat{\section}{\large\bfseries}{\thesection.}{0.5em}{}
\titleformat{\subsection}{\normalsize\bfseries}{\thesubsection.}{0.5em}{}
\titleformat{\subsubsection}{\normalsize\itshape}{\thesubsubsection.}{0.5em}{}

\title{\textbf{Inverse Materials Design via Joint Generation of Crystal Structures and Local Electronic Descriptors}}

\author[1]{Ibuki Okuda}
\author[1]{Izumi Takahara}
\author[1,2]{Teruyasu Mizoguchi}

\affil[1]{Department of Materials Engineering, The University of Tokyo}
\affil[2]{Institute of Industrial Science, The University of Tokyo}
\date{} 

\begin{document}

\twocolumn[
\begin{@twocolumnfalse}
    \maketitle

    \begin{abstract}
      Inverse design of inorganic crystals, in which structures are generated to satisfy a target property while preserving diversity and physical plausibility, remains more demanding than \textit{ab initio} generation, as property conditioning can degrade the structural quality that current generative models otherwise achieve. We propose a diffusion framework that jointly denoises crystal-structure variables and site-resolved local electronic descriptors through a shared score network. As representative descriptors, we adopt Bader charge and atomic density of states (atomic DOS). Under both band-gap- and formation-energy-conditioned generation, the joint models achieved higher success rates than the structure-only baseline in most target conditions, while simultaneously increasing the fraction of generated structures that satisfy uniqueness, novelty, thermodynamic stability, and physical validity (VSUN criteria). A dummy-variable control confirms that these gains originate from the electronic content of the descriptors rather than from auxiliary site-wise variables. The generated Bader charges agree with DFT references with an MAE of $5.5 \times 10^{-2}$~e on stable structures, and the generated atomic DOS captures the coarse spectral profile of the DFT reference around the modal accuracy range, although finer details and accuracy vary with elemental species. These results establish local electronic descriptors as effective generative variables that serve two complementary roles: broadening the explored materials space through increased structural diversity, and mitigating the trade-off between property targeting and structural quality by guiding the structural trajectory toward electronically plausible configurations during joint denoising. 
    \end{abstract}
    \vspace{1em}
    \noindent\textbf{Keywords:} crystal structure generation; diffusion model; inverse materials design; local electronic descriptor; Bader charge; density of states
    
    \vspace{1.5em}
\end{@twocolumnfalse}
]

\section{Introduction}

\begin{figure*}[t]
  \centering
  \includegraphics[width=\textwidth]{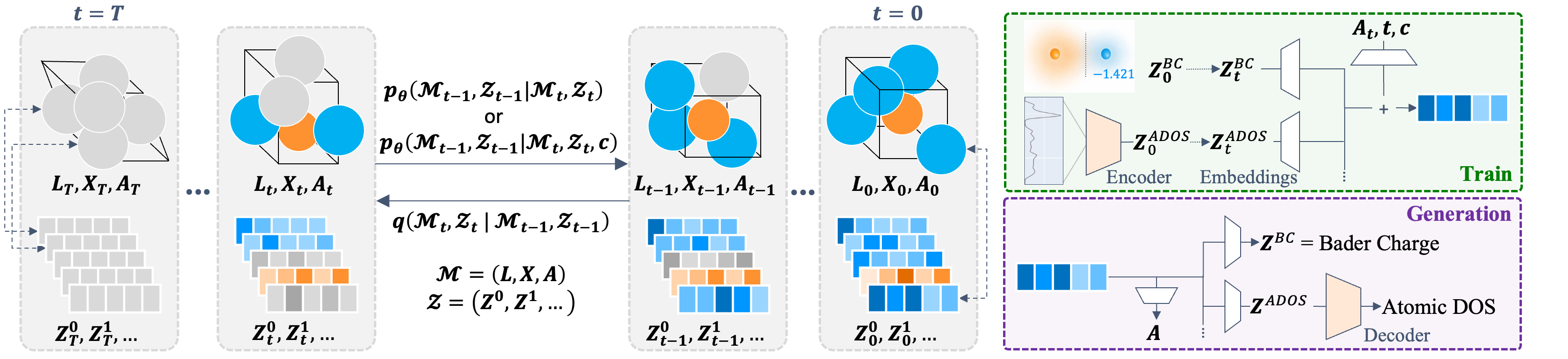}
  \caption{
    Schematic illustration of the proposed diffusion framework for the joint generation of crystal structures $\mathcal{M}$ and local electronic descriptors $\mathcal{Z}$. In addition to the structural variables treated in crystal generation, namely the lattice $\bm{L}$, atomic positions $\bm{X}$ and atom types $\bm{A}$, the model simultaneously denoises and generates site-resolved local electronic quantities $\bm{Z}^0,\bm{Z}^1,\ldots$. In this study, Bader charge and atomic DOS latent variables are compared as representative local electronic descriptors, and the right panels illustrate how each is incorporated into the framework. During training (green panel), the noised descriptors $\bm{Z}^{\mathrm{BC}}_t$ and $\bm{Z}^{\mathrm{ADOS}}_t$ are embedded together with $\bm{A}_t$, $t$, and the optional conditioning $c$ into the node features of the score network, where $\bm{Z}^{\mathrm{ADOS}}$ is obtained by compressing the atomic DOS spectrum with a pretrained encoder. At generation (purple panel), the denoised node features yield the atom type $\bm{A}$ together with the local electronic descriptors $\bm{Z}^{\mathrm{BC}}$ and $\bm{Z}^{\mathrm{ADOS}}$ through dedicated regression heads, and the generated $\bm{Z}^{\mathrm{ADOS}}$ is decoded back to the atomic DOS spectrum.
  }
  \label{fig:model_overview}
\end{figure*}

Generative models for inorganic crystals have advanced rapidly in recent years and are increasingly regarded as a promising route to materials discovery beyond conventional screening-based workflows \cite{sanchez2018inverse,lu2022inverse,debreuck2025review,merchant2023scaling}. Instead of enumerating candidate compositions and structures and subsequently evaluating their properties, generative models aim to learn the distribution of crystal structures and to sample new candidates directly from it. This paradigm is particularly attractive in materials science because the space of plausible compounds and periodic structures is vast, whereas first-principles evaluation remains computationally expensive even in high-throughput settings \cite{curtarolo2012aflow}.

Within this context, \textit{ab initio} crystal generation has emerged as a promising research direction. A variety of generative approaches have been proposed, including variational-autoencoder (VAE)-based models \cite{noh2019inverse,court2020vae,ren2022invertible}, GAN-based models \cite{nouira2018crystalgan,kim2020gan}, flow-matching-based models \cite{miller2024flowmm}, and diffusion-based models \cite{takahara2024generative,luo2024deep,xie2022cdvae,jiao2023diffcsp,jiao2024diffcsppp,levy2025symmcd,zeni2025mattergen}, and recent studies have begun to benchmark their performance systematically \cite{betala2025lematgenbench,debreuck2025review}. Among them, diffusion models have shown especially strong performance. CDVAE introduced a diffusion-based decoder for periodic crystals with a physically motivated inductive bias \cite{xie2022cdvae}; DiffCSP showed that jointly modeling lattice variables and atomic coordinates is highly effective for crystal generation \cite{jiao2023diffcsp}; DiffCSP++ \cite{jiao2024diffcsppp} and SymmCD \cite{levy2025symmcd} improved crystal generation by explicitly incorporating space-group symmetry; and MatterGen further demonstrated strong performance in generating stable, diverse, and novel inorganic materials, while also enabling property-conditioned generation through fine-tuning \cite{zeni2025mattergen}. These advances establish generative modeling as a powerful framework for crystal discovery.

While generative models for \textit{ab initio} crystal generation have advanced rapidly in recent years, they remain insufficient for practical materials applications, where the objective shifts toward inverse design: generating crystal structures that satisfy a desired target property, such as a specific band gap or mechanical response, while remaining physically plausible and diverse \cite{zunger2018inverse,lu2022inverse,park2024walsh_genai}. In this setting, success is not determined solely by property matching. A useful inverse-design model must also preserve the diversity and physical plausibility of the generated structures, because a narrow or unrealistic candidate set substantially limits downstream applications. In other words, inverse design is more demanding than \textit{ab initio} generation: the model must satisfy the target property while avoiding collapse to a small subset of memorized or marginally valid structures. This additional demand is often non-trivial in practice, as the imposed conditioning can erode the validity and diversity that current generative models attain in the \textit{ab initio} setting. Improving inverse-design performance without such degradation therefore remains a central challenge. A possible route to address this challenge lies in reconsidering what information the generative model should operate on. 

A key observation motivating this work is that materials properties are often closely linked to local electronic states \cite{deng2023chgnet,xie2018cgcnn,chen2019megnet}. Existing crystal generative models primarily describe materials through structural variables such as lattice parameters, atomic species, and atomic positions\cite{xie2022cdvae,jiao2023diffcsp,zeni2025mattergen}. However, the properties and physical realizability of a crystal are sometimes governed by its local electronic structure, which emerges from the underlying crystal structure. While this dependence is implicitly encoded in structural representations, local electronic states, such as charge redistribution and site-resolved spectral features, provide a more explicit description of chemical bonding, stability, and target properties. This suggests that explicitly modeling such local electronic information can provide a more effective inductive bias for inverse design, particularly when targeting electronically driven properties.

Recent work has started to incorporate electronic information directly into crystal generation. For example, ChargeDIFF introduced a diffusion-based framework that augments crystal generation with three-dimensional charge density, demonstrating that explicit electronic information can improve inorganic crystal generation \cite{park2025chargediff}. This result clearly highlights the importance of electronic structure as a generative modality. At the same time, charge density is a highly detailed and high-dimensional representation and its physical interpretation at the site level is often nontrivial. Although expressive, charge density is not necessarily the most direct representation for learning the site-level electronic factors that are most relevant to property targeting, structural validity, and diversity in inverse design.

In this study, we propose MatterGen-$e^-$, a diffusion-based framework that jointly generates crystal-structure variables and interpretable local electronic descriptors (Fig.~\ref{fig:model_overview}), built upon the architecture of MatterGen \cite{zeni2025mattergen}. Rather than treating electronic quantities as post hoc predictions made after structure generation, the model incorporates site-resolved electronic variables directly into the generative process so that structure and local electronic states evolve jointly during denoising. In this way, the model is encouraged to learn node representations that are not only geometrically consistent but also electronically plausible at the atomic level. Our central hypothesis is that such joint learning provides a more effective inductive bias for inverse design: if generated structures are accompanied by locally reasonable electronic states, the model can better capture structure--property relationships while preserving the validity and diversity of generated crystals.

As representative local electronic descriptors, we consider two distinct representations of local electronic structure: Bader charge and atom-level electronic density of states (atomic DOS). Bader charge is a site-resolved scalar quantity obtained by partitioning the electron density into atomic basins and integrating the electron density within each atomic basin \cite{bader1990atoms,henkelman2006fast,sanville2007improved,tang2009grid,yu2011accurate}. Because it reflects both elemental identity and local bonding environment, it provides a compact description of the local electronic state. Atomic DOS, in contrast, provides a richer site-resolved spectral description of the local electronic structure. By studying both descriptors, we examine the proposed framework at two qualitatively different levels of local electronic representation: a compact scalar quantity and a higher-dimensional spectral quantity.

Building on this formulation, we investigate how explicitly modeling local, site-resolved electronic descriptors within a crystal generative framework influences inverse design, moving beyond approaches that rely solely on structural variables. By jointly generating these descriptors alongside crystal structures, the proposed framework improves inverse-design performance while maintaining, and in some cases enhancing, the validity and diversity of generated structures. These results demonstrate that local electronic information provides an effective inductive bias for physically plausible and property-targeted generation.

\section{Results}

\begin{figure*}[t]
  \centering
  \includegraphics[width=\textwidth]{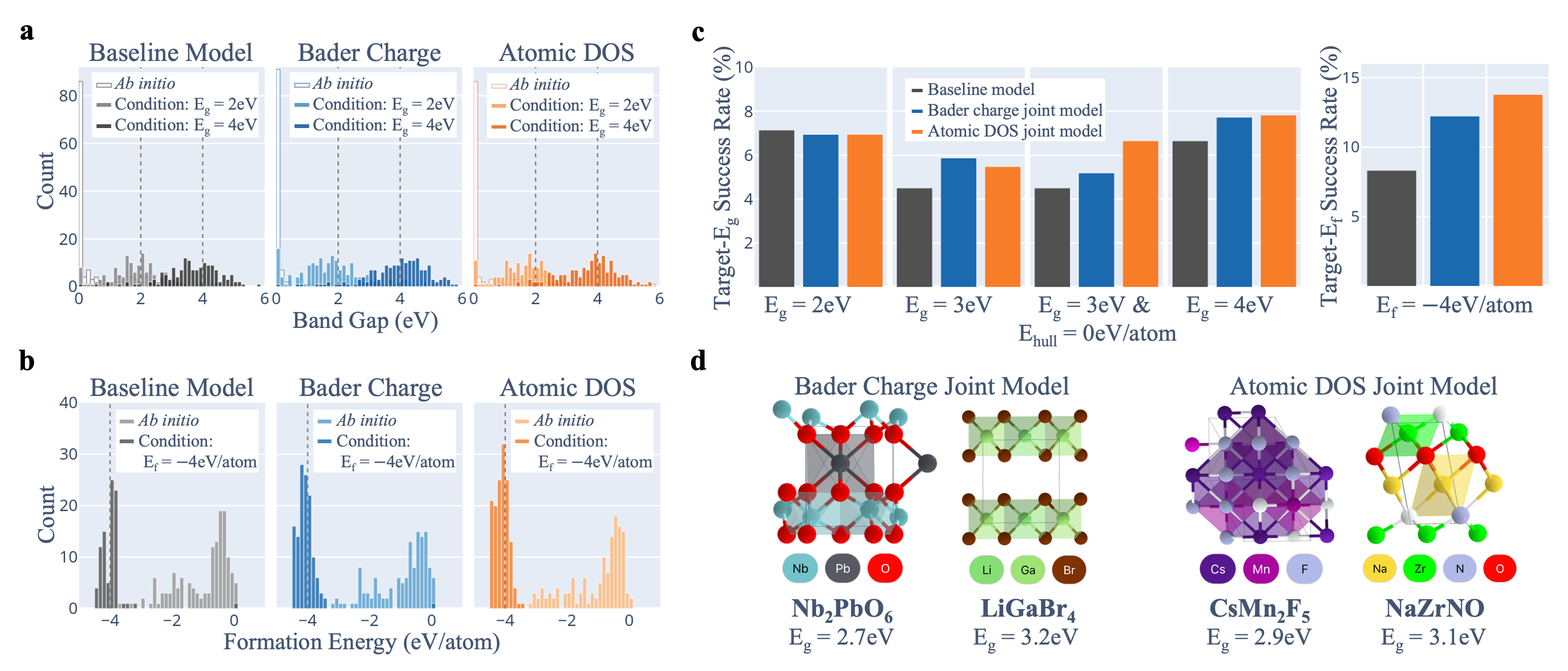}
  \caption{
    Band-gap and formation-energy distributions and conditional generation success rates for VSUN structures generated by the baseline (gray), Bader charge joint (blue), and atomic DOS joint (orange) models.
    \textbf{a.}~DFT band-gap distributions for \textit{ab initio} generation and conditional generation with targets $E_\mathrm{g} = 2$ and $4$~eV. Dashed vertical lines indicate the respective target values.
    \textbf{b.}~DFT formation-energy distributions for \textit{ab initio} generation and conditional generation with target $E_\mathrm{f} = -4$~eV/atom.
    \textbf{c.}~Success rates for each conditional generation target. The four left bar charts show the target-$E_\mathrm{g}$ success rates for conditional generation with $E_\mathrm{g} = 2$~eV, $E_\mathrm{g} = 3$~eV, $E_\mathrm{g} = 3$~eV \& $E_\mathrm{hull} = 0$~eV/atom, and $E_\mathrm{g} = 4$~eV. The right bar chart shows the target-$E_\mathrm{f}$ success rate for conditional generation with $E_\mathrm{f} = -4$~eV/atom.
    \textbf{d.}~Representative VSUN structures generated under the $E_\mathrm{g} = 3$~eV condition by the Bader charge joint model and the atomic DOS joint model.
  }
  \label{main_results}
\end{figure*}

\begin{table*}[t]
  \centering
  \caption{Comparison of fractions of generated structures satisfying the combined criteria of V/S/U/N. Fractions are relative to the total number of generated structures ($1024$) in each setting. Boldface indicates values whose corresponding proportions differ significantly from those of the baseline model under the same condition, based on a two-sided two-proportion z-test at p < 0.05.}
  \label{tab:static_property_change}
  \small
  \setlength{\tabcolsep}{5pt}
  \begin{tabular}{llccc}
    \toprule
    \multirow{2}{*}[-0.8ex]{Condition} & \multirow{2}{*}[-0.8ex]{Model} & \multicolumn{3}{c}{Fraction} \\
    \cmidrule(lr){3-5}
      &  & UN & SUN & VSUN \\
    \midrule
    \multirow{3}{*}{\textit{ab initio}}
      & Baseline & 68.1\% & 27.7\% & 14.0\% \\
      & Bader charge& \textbf{75.3}\% & 28.0\% & 13.3\% \\
      & Atomic DOS & 69.9\% & 26.3\% & 13.9\% \\
    \midrule
    \multirow{3}{*}{$E_\mathrm{g}=2$ eV}
      & Baseline & 68.5\% & 27.1\% & 14.0\% \\
      & Bader charge & \textbf{83.2\%} & \textbf{32.7\%} & \textbf{18.5\%} \\
      & Atomic DOS & \textbf{75.8\%} & \textbf{33.5\%} & 14.5\% \\
    \midrule
    \multirow{3}{*}{$E_\mathrm{g}=3$ eV}
      & Baseline & 68.4\% & 22.6\% & 11.4\% \\
      & Bader charge & \textbf{80.3\%} & \textbf{28.4\%} & \textbf{15.9\%} \\
      & Atomic DOS & 72.2\% & \textbf{26.9\%} & \textbf{15.4\%} \\
    \midrule
    \multirow{3}{*}[0pt]{\raggedright\arraybackslash\shortstack[l]{$E_\mathrm{g}=3$ eV\\\& $E_\mathrm{hull}=0$ eV/atom}}
    & Baseline & 65.6\% & 22.8\% & 11.4\% \\
    & Bader charge & \textbf{78.9\%} & \textbf{30.4\%} & \textbf{17.5\%} \\
    & Atomic DOS   & \textbf{73.2\%} & \textbf{31.3\%} & \textbf{16.5\%} \\
    \midrule
    \multirow{3}{*}{$E_\mathrm{g}=4$ eV}
    & Baseline & 67.6\% & 24.3\% & 14.7\% \\
    & Bader charge & \textbf{79.1\%} & 28.1\% & \textbf{17.9\%} \\
    & Atomic DOS   & \textbf{74.8\%} & 27.6\% & 16.7\% \\
    \midrule
    \multirow{3}{*}{$E_\mathrm{f}=-4$ eV/atom}
      & Baseline & 60.8\% & 18.9\% & 8.6\% \\
      & Bader charge & \textbf{69.6\%} & 18.8\% & \textbf{12.5\%} \\
      & Atomic DOS   & \textbf{70.0\%} & \textbf{27.1\%} & \textbf{13.9\%} \\  
    \bottomrule
  \end{tabular}
\end{table*}

\subsection{Property-Conditioned Generation Performance}

We evaluated the performance of property-conditioned crystal generation using three models: a baseline model that generates only crystal structures, a joint model with Bader charge generation, and a joint model with atomic DOS generation. In each setting, the model generated 1024 candidate structures, allowing for statistical comparison of the resulting structural distributions. The generated structures were then evaluated in a stepwise manner. First, uniqueness (U) and novelty (N) were assessed, and structures satisfying both criteria were identified as UN structures. Next, DFT calculations were performed for the UN structures to obtain relaxed structures and electronic states, from which stability (S) was evaluated to define the SUN subset. In addition, validity (V) was assessed based on compositional criteria. Combining these criteria yields the VSUN structures, whose target-property distributions are analyzed and compared with the imposed conditional values. This framework enables evaluation of the models not only in terms of property matching, but also in terms of the physical plausibility and novelty of the generated materials.

As a first application of this evaluation procedure, we examined the results of band-gap-conditioned generation for target values of $E_\mathrm{g}=2$, $3$, and $4$~eV. For each target value and each model, the actual band gaps of the resulting VSUN structures were calculated and compared across models (Fig.~\ref{main_results}\textbf{a}). To quantify the degree of concentration for the target value, we defined the success rate as the fraction of generated VSUN structures, out of all 1024 samples, whose calculated band gaps deviate by no more than  $\pm 0.5$~eV of the target value. For the $E_\mathrm{g}=2$~eV condition, the success rates were 7.1\% for the baseline model, 6.9\% for the Bader-charge joint model, and 6.5\% for the atomic-DOS joint model. For the $E_\mathrm{g}=3$~eV condition, the corresponding values were 4.5\%, 5.9\%, and 5.5\%. For the $E_\mathrm{g}=4$~eV condition, the success rates were 6.6\%, 7.7\%, and 7.8\%, respectively (Fig.~\ref{main_results}\textbf{c}). Overall, joint generation of local electronic descriptors showed improved band-gap-conditioned success rates for selected target values, with clear gains at $E_\mathrm{g}=3$ and $4$~eV.

A further contrast emerged when comparing the results of the $E_\mathrm{g}=3$~eV condition with those of the joint condition imposing both $E_\mathrm{g}=3$~eV and $E_\mathrm{hull}=0$~eV/atom. This comparison addresses a practically important question: whether the model can simultaneously satisfy a band-gap target and thermodynamic stability, two constraints that may be difficult to fulfill jointly. In the Bader-charge joint model, adding the $E_\mathrm{hull}=0$~eV/atom condition broadened the band-gap distribution, and the corresponding success rate decreased slightly to 5.2\%. By contrast, in the atomic-DOS joint model, the additional $E_\mathrm{hull}=0$~eV/atom condition led to a better concentration of the distribution around the target value, and the success rate increased to 6.6\% (Fig.~\ref{main_results}\textbf{c}). One possible explanation for this improvement is that the additional stability-related condition of $E_\mathrm{hull}=0$~eV/atom affects the two joint models in different ways. In the Bader-charge joint model, the added condition appears to shift the generation process toward structures that better satisfy stability-related criteria, resulting in an increased number of VSUN structures but a weaker concentration around the target band gap. In the atomic-DOS joint model, by contrast, the additional condition appears to promote generation that better balances structural stability with band-gap targeting, leading to a distribution that remains relatively concentrated near the target value while also improving the survival of generated structures through the screening process. This interpretation is also consistent with the subsequent comparison of the numbers of VSUN-qualified structures described later, where the different roles of the two descriptors become more apparent.

Having established the behavior of the joint models under band-gap conditioning, we next turn to a qualitatively different target property to further probe the generalization of the proposed framework. We next examined the results of formation-energy-conditioned generation with a target value of $E_\mathrm{f} = -4$~eV/atom. Structures with formation energies below $-4$~eV/atom are exceedingly rare in the training dataset, comprising 0.4\% of all entries, and they appear only sporadically in \textit{ab initio} generation. Despite this rarity, joint generation of local electronic descriptors led to a marked increase in the number of VSUN structures with formation energies concentrated near the target value (Fig.~\ref{main_results}\textbf{b}). Quantitatively, the formation-energy success rates were 8.3\%, 12.2\%, and 13.8\% for the baseline, Bader-charge joint, and atomic-DOS joint models, respectively (Fig.~\ref{main_results}\textbf{c}). The improvement was substantially more pronounced than that observed in the band-gap-conditioned case. This result indicates that joint generation of site-resolved electronic quantities contributes substantially to the model's ability to learn the thermodynamic stability landscape of crystal structures, enabling targeted generation of highly stable materials that would otherwise be virtually difficult to access through \textit{ab initio} or structure-only conditional generation.

\subsection{Effects on Structural Plausibility, Stability, Uniqueness and Novelty}

\begin{figure}[t]
  \centering
  \includegraphics[width=\columnwidth]{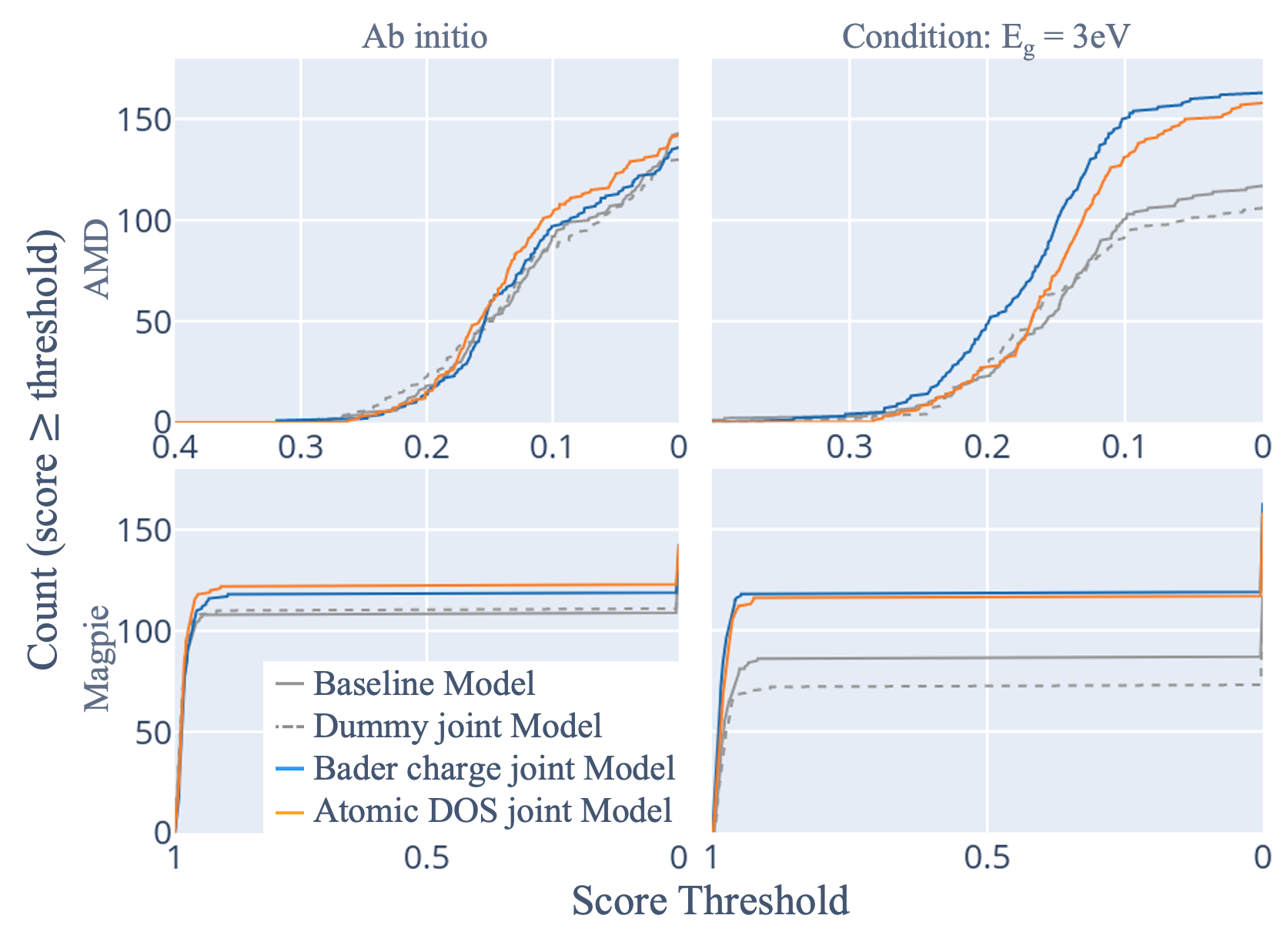}
  \caption{
    Novelty of VSUN structures generated by the baseline (gray solid), dummy joint (gray dashed), Bader charge joint (blue), and atomic DOS joint (orange) models.
    Each panel shows the number of generated structures whose novelty score exceeds a given threshold, evaluated for \textit{ab initio} generation (left column) and conditional generation with target $E_\mathrm{g} = 3$~eV (right column), using the structure-based AMD metric (top row) and the composition-based Magpie metric (bottom row).
  }
  \label{fig:novelty}
\end{figure}

Table~\ref{tab:static_property_change} presents the UN, SUN, and VSUN counts under property-conditioned generation, demonstrating that the joint models consistently outperform the baseline across almost all conditions examined, with most improvements statistically significant under a two-sided two-proportion z-test at $p<0.05$ (boldface entries).  Notably, this improvement contrasts with the behavior of the baseline model, for which imposing a property condition generally reduced these counts relative to the unconditional case, providing clear evidence of a trade-off between property targeting and structural quality. By contrast, the joint models largely preserved or even improved these counts under the same conditions, indicating that joint generation of local electronic variables enables the model to satisfy the target property without compromising thermodynamic stability or physical validity.

Among the individual metrics, the enhancement was particularly pronounced for the UN count, where the Bader-charge joint model exhibited a larger improvement than the atomic DOS joint model across most conditional settings. A further notable trend emerged when comparing the $E_\mathrm{g}=3$~eV condition with the joint condition imposing both $E_\mathrm{g}=3$~eV and $E_\mathrm{hull}=0$~eV/atom: the addition of the explicit stability constraint led to an increase in the SUN and VSUN counts for both joint models, confirming that the stability condition indeed raises the fraction of thermodynamically stable structures among the generated candidates.

Under \textit{ab initio} generation, the differences between the baseline the joint models in these binary novelty metrics were minimal and largely inconclusive (Table~\ref{tab:static_property_change}). To obtain a more fine-grained view of the generated materials space, we next examine continuous novelty metrics \cite{negishi2025continuousnovelty}. As shown in Fig.~\ref{fig:novelty}, the joint models produce VSUN structures with systematically higher novelty scores than the baseline, with the clearest improvement observed for the composition-based Magpie metric \cite{ward2016magpie}. These results indicate that, even under \textit{ab initio} generation, the proposed framework effectively expands the explored materials space, particularly along the compositional axis.

To verify that these benefits genuinely originate from the electronic content of the site-resolved descriptors, rather than from the mere presence of additional site-wise variables during training, we performed a control experiment using a dummy-variable joint model. In this model, each atomic site was assigned a physically meaningless and randomly generated scalar variable, and the diffusion model was trained to jointly generate these dummy variables together with the structural variables in exactly the same manner as the Bader-charge and atomic-DOS joint models. We generated 1024 structures for each setting and evaluated the results; the fractions of UN, SUN, and VSUN structures obtained with the dummy-variable model were 68.8\%, 22.6\%, and 12.7\% for \textit{ab initio} generation, and 64.3\%, 19.3\%, and 10.4\% for the $E_\mathrm{g}=3$~eV conditional generation, respectively, which are values comparable to or lower than those of the baseline, with no sign of the improvement observed for the electronic-descriptor joint models. Continuous novelty analysis further confirmed this trend (Fig.~\ref{fig:novelty}). These results demonstrate that the improvements cannot be attributed to auxiliary variables per se, but require the jointly generated quantities to genuinely reflect the local electronic structure at each atomic site. These results indicate that the effectiveness of expanding the generative space depends not on its dimensionality, but on the physical meaningfulness of the additional variables. Physically informed descriptors enable the model to capture richer structure-property relationships, whereas uninformative variables do not provide the same benefit.

\subsection{Accuracy of Bader Charge Generation}

\begin{figure}[t]
  \centering
  \includegraphics[width=\columnwidth]{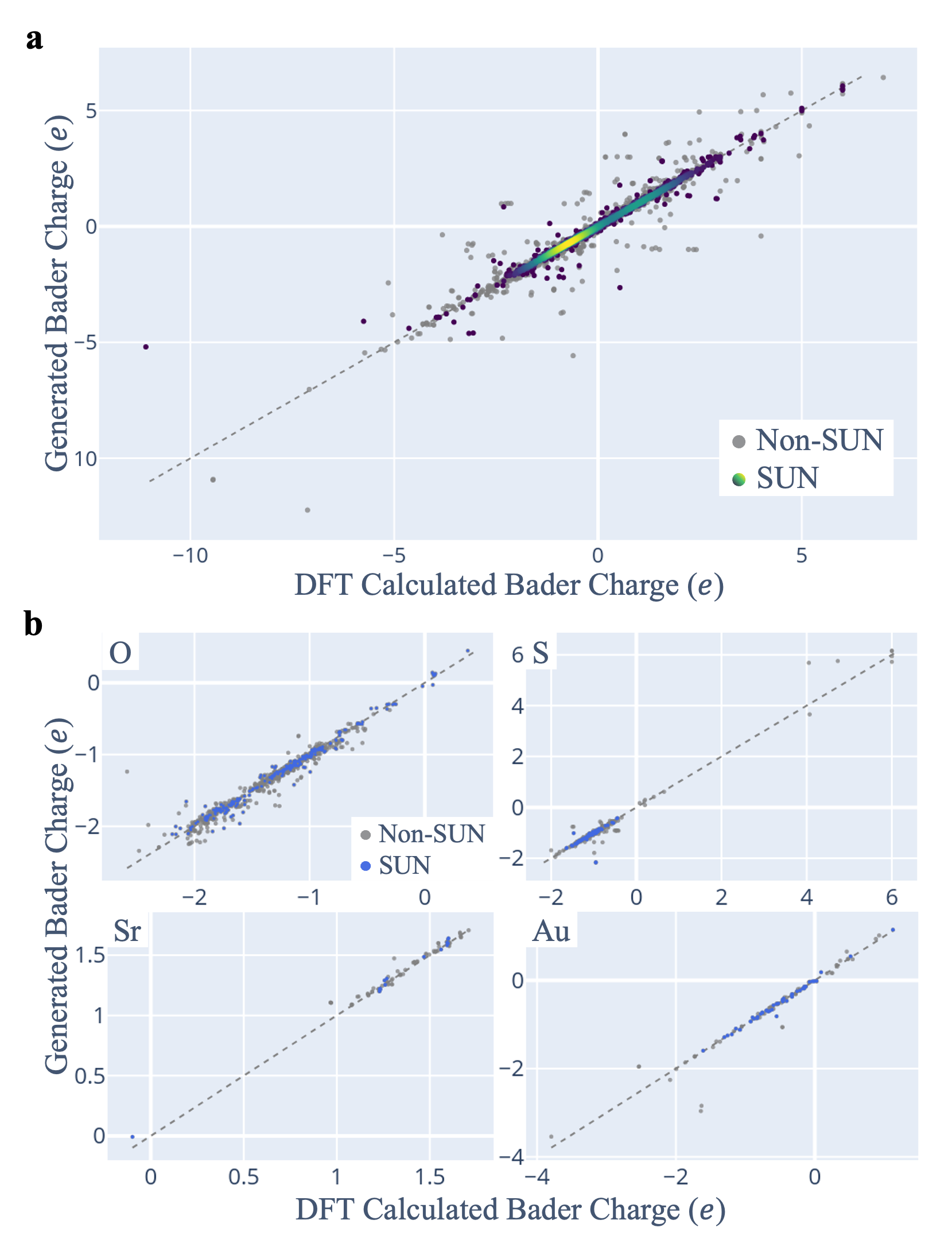}
  \caption{
    Comparison between generated and DFT-calculated Bader charges for atoms in the generated structures.
    \textbf{a.}~Overall correlation for all atoms across all generated structures. Atoms belonging to SUN structures (MAE: $5.5 \times 10^{-2}$~$e$) are shown using a density colormap, and those from non-SUN structures are plotted as gray points.
    \textbf{b.}~Element-resolved correlations for O, S, Sr, and Au. In each panel, atoms from SUN structures (blue) and non-SUN structures (gray) are shown separately.
  }
  \label{fig:bader_charge}
\end{figure}

While joint generation of local electronic descriptors was found to improve the quality of the generated structures, it is also important to verify whether the generated descriptors are physically consistent with their corresponding structures. To this end, we assess the accuracy of the generated Bader charges for structures obtained under \textit{ab initio} generation. Specifically, the jointly generated Bader charges were compared with DFT reference values computed after structural relaxation of each structure. This accuracy was assessed both over all generated structures and over the subset of SUN structures.

Overall, the generated Bader charges agreed well with the DFT-calculated reference values, yielding a mean absolute error (MAE) of $7.2 \times 10^{-2}$~e over all generated structures, as shown in Fig.~\ref{fig:bader_charge}\textbf{a}. When the analysis was restricted to SUN structures, the MAE decreased to $5.5 \times 10^{-2}$~e. Consistent with this trend, large deviations from the DFT reference are observed only for a limited number of points in the full set, whereas such deviations are rarely found for the SUN subset, whose generated charges remain concentrated around the reference relation. This result indicates that Bader charge generation is more accurate for structures that satisfy the SUN criterion.

Examining the results by element, the generated charges follow the overall trend of the DFT-calculated values for each element, as exemplified by the results for O, S, Sr, and Au shown in Fig.~\ref{fig:bader_charge}\textbf{b}. Notably, even in regions corresponding to relatively rare or extreme charge states, the generated values remain aligned with the reference trend. Element-wise evaluation further showed that the MAE is smaller than 0.1~e for most elements (Fig.~\ref{fig:accuracy_periodic}\textbf{b}), indicating that the model captures not only the average charge behavior but also the element-specific variation in local charge states with good quantitative accuracy. Taken together, these results demonstrate that the model has successfully learned the correlation between crystal structure and the corresponding Bader-charge distribution.

\subsection{Accuracy of Atomic DOS Generation}

\begin{figure*}[t]
  \centering
  \includegraphics[width=\textwidth]{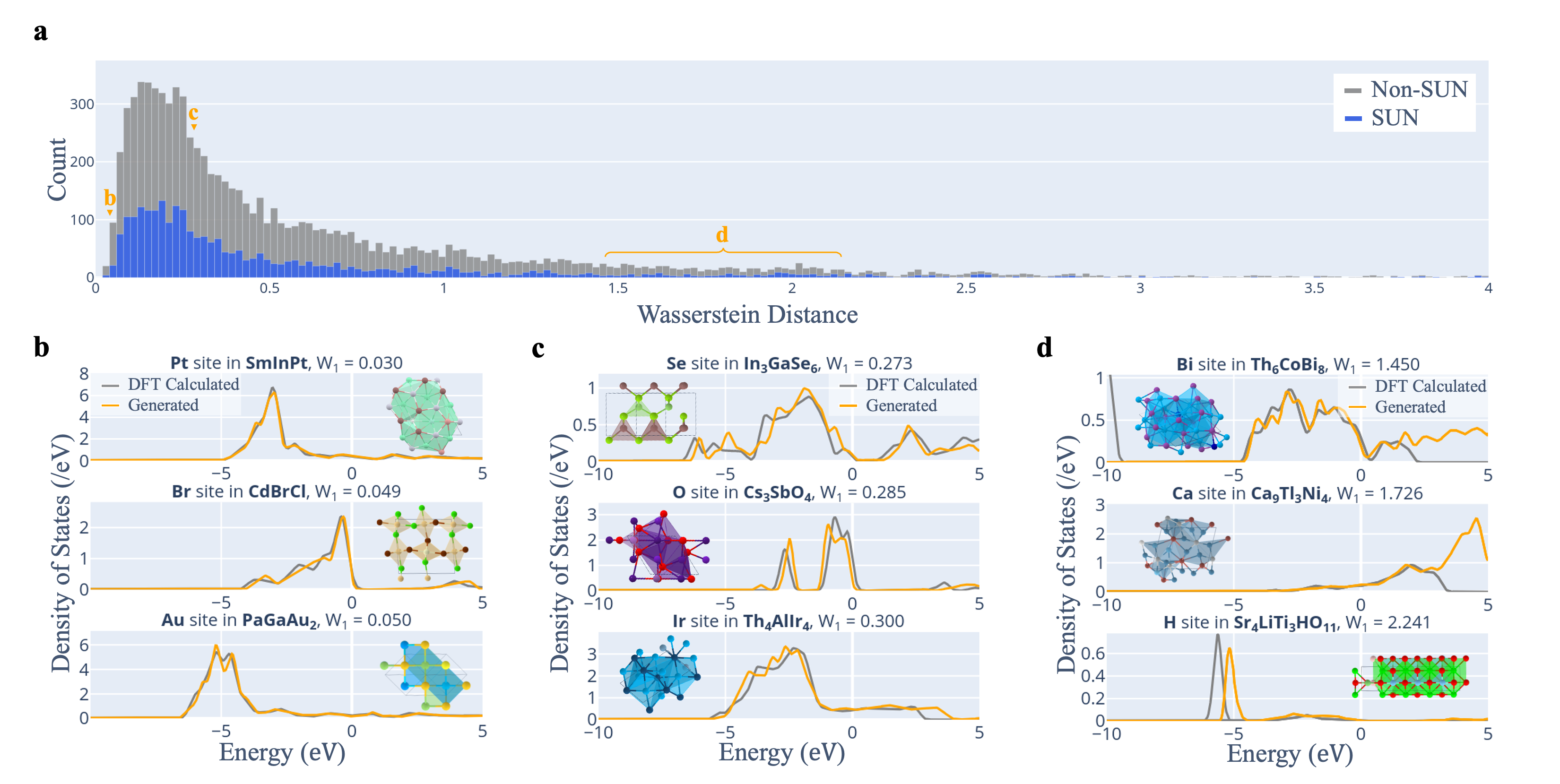}
  \caption{
    Evaluation of atomic DOS generation quality using the Wasserstein distance $W_1$ between generated and DFT-calculated atomic DOS profiles.
    \textbf{a.}~Distribution of $W_1$ for all generated structures (gray) and SUN structures (blue). Annotations indicate the $W_1$ ranges corresponding to the representative examples shown in \textbf{b}, \textbf{c}, and \textbf{d}.
    \textbf{b.}~Representative generated (orange) and DFT-calculated (gray) atomic DOS profiles for three sites with small $W_1$ ($\leq 0.06$; below the 1st percentile among SUN structures).
    \textbf{c.}~Same as \textbf{b}, but for three sites with modal $W_1$ ($0.273$--$0.31$; 45th--50th percentile).
    \textbf{d.}~Same as \textbf{b}, but for three sites with large $W_1$ ($\geq 1.43$; above the 90th percentile).
    Crystal structures are shown as insets in each panel.
    }
  \label{fig:atomic_dos}
\end{figure*}

We next examine the accuracy of the generated atomic DOS. For the atomic-DOS joint model, we evaluated the accuracy of the jointly generated atomic DOS by comparing it with DFT reference DOS computed for the corresponding structures obtained under \textit{ab initio} generation. Prior to comparison, both the generated and DFT calculated DOS were normalized such that their integrated intensity equals unity, and the difference in spectral shape between the generated and the ground truth was quantified using the Wasserstein distance $W_1$. Because this metric captures discrepancies between overall distributions rather than pointwise deviations, it is well suited to assessing DOS-profile similarity even when peak positions and relative intensities differ.

Figure~\ref{fig:atomic_dos}\textbf{a} shows the histogram of Wasserstein distances for the generated structures, revealing a broad distribution that indicates substantial variation in generation accuracy across samples. The histogram for the SUN subset shows a slight improvement in summary statistics such as the mean and median relative to the full set, while the overall shape of the distribution remains largely unchanged.

Representative DOS pairs are also shown for three cases: small (below the 1st percentile among SUN structures, $W_1 \leq 0.06$), modal (45th--50th percentile, $0.27 \leq W_1 \leq 0.31$), and large (above the 90th percentile, $1.43 \leq W_1$) Wasserstein distances (Fig.~\ref{fig:atomic_dos}\textbf{b},\textbf{c},\textbf{d}). For the smallest distances, the generated DOS is in near-complete agreement with the DFT reference, whereas little correspondence is found in the largest-distance cases. For samples with distances near the most frequent range, differences remain in finer details such as peak position and relative peak intensity; however, the overall spectral profile is still reasonably reproduced. Thus, even around the modal range of the distribution, the model is able to capture the coarse spectral features of the atomic DOS.

Through the investigation of the generated accuracy of atomic DOS, we found that generation accuracy varied substantially across elemental species, revealing that the learnability of atomic DOS is strongly governed by chemical identity rather than by structural complexity alone. Some elements were associated with comparatively small MAE of Wasserstein distance, whereas others showed much larger, indicating that the learnability of atomic DOS depends sensitively on chemical identity (Fig.~\ref{fig:accuracy_periodic}\textbf{b}).
In particular, transition-metal elements (with $3d$, $4d$, and $5d$ valence orbitals) tended to yield relatively accurate DOS generation (Fig.~\ref{fig:accuracy_periodic}\textbf{b}). This behavior may be attributed to the fact that bands derived from $d$ orbitals often show less variation in spectral shape and a narrower spread in characteristic peak energies across different local environments. Such regularity is likely to make these features more amenable to statistical learning. By contrast, relatively high MAE was observed for light nonmetallic elements such as B, C, and N (Fig.~\ref{fig:accuracy_periodic}\textbf{b}). A possible explanation is the greater diversity of electronic states associated with these elements, including bands originating from $sp$-hybridized orbitals and ionicity-related bonding states. Because these states are strongly influenced by the local bonding environment and can vary substantially with crystal structure, the corresponding atomic DOS exhibits a wide variety of spectral patterns, making accurate generation inherently more difficult. Interestingly, this trend is in a sense the inverse of the difficulty encountered in DFT calculations, where transition-metal elements are notoriously challenging due to strong electron correlation effects in some oxides materials, whereas light nonmetallic elements are comparatively straightforward. Here, the situation is reversed: the spectral regularity of $d$-orbital-derived bands makes transition metals easier to learn, while the environmental sensitivity of  $sp$-hybridized states makes light elements the harder cases for generative modeling.

The results indicate that the quality of atomic DOS generation varies considerably with both the crystal structure and elemental species. Although the model is able to reproduce the overall tendency of the DOS in many cases, the attainable accuracy remains moderate because of the large diversity of possible local electronic structures.

\section{Discussion}

The central finding of this study is that jointly generating local electronic descriptors with crystal structures improves the generation of physically plausible novel materials, which in turn leads to higher success rates of property-conditioned generation. The results suggest that incorporating local electronic information into the generative process improves robustness in inverse design, enabling property conditioning without the substantial loss of structural quality observed in the structure-only baseline.

A notable feature of the present results is that the clearest improvement first appears in the UN fraction. This trend can be interpreted in terms of the stochasticity of ancestral sampling in diffusion models. In the reverse process, the state at step $t-1$ is sampled from a conditional distribution defined by the state at step $t$. In the joint model, the local electronic descriptors are also sampled through this process and therefore exhibit stepwise variation during denoising. Because the shared score network predicts structural updates conditioned on the current noisy electronic state as well as the structural variables, the structural trajectory is guided toward configurations consistent with multiple plausible local electronic states rather than following a purely geometric denoising path. This provides a plausible explanation for the increase in structural diversity. This view is also consistent with the strongest UN improvement observed in the Bader-charge joint model and with the absence of comparable diversity gains in the dummy-variable joint model.

Another key findings lies in the contrast between \textit{ab initio} and property-conditioned generation. For the baseline structure-only model, imposing the band-gap condition reduced the SUN and VSUN fractions relative to \textit{ab initio} generation. By contrast, for the models with joint generation of local electronic descriptors, these fractions were largely maintained and in some cases slightly improved. A plausible explanation for this contrast is that local-electronic joint generation mitigates the trade-off between property targeting and structural quality that emerges in structure-only conditional generation. In the baseline model, property conditioning acts only through correlations between the target property and structural geometry learned in the latent manifold, so the sampling trajectory may be driven toward structures that better satisfy the target condition at the expense of stability, validity, or diversity. By contrast, in the joint models, the reverse process is guided not only by the target property but also by site-resolved local electronic descriptors related to bonding and local chemical environment. This additional guidance may help maintain electronic and chemical plausibility during sampling, thereby allowing the model to satisfy the target condition while maintaining physically reasonable structures.

The results across multiple conditional-generation settings allow further interpretation of how Bader charge and atomic DOS joint generation contribute differently to the overall improvements. Bader charge carries information primarily related to basic physical plausibility, such as local charge balance and bonding environment, and its joint generation therefore appears to contribute mainly to raising the fraction of structures that satisfy validity and stability criteria, resulting in a comparatively high VSUN yield across conditions. Atomic DOS, by contrast, encodes richer spectral information more directly related to material properties such as the band gap, and its joint generation therefore tends to contribute more specifically to the success rate of property-conditioned generation among VSUN structures. This distinction reflects the difference in information content between the two descriptors: Bader charge provides compact but physically grounded constraints on local charge balance, while atomic DOS encodes richer spectral features more directly tied to electronic properties.  This interpretation is broadly consistent with, for example, the observation that the atomic-DOS joint model better maintains band-gap targeting under the additional $E_\mathrm{hull}=0$~eV/atom constraint, whereas the Bader-charge joint model shows a comparatively larger increase in absolute VSUN counts.

However, these roles should not be regarded as sharply separated. A contributing factor is that atomic DOS generation has not reached complete accuracy, as evidenced by the moderate overall Wasserstein distances and the strong element dependence reported above. Because the model operates on latent variables of a compressed DOS representation, some physically meaningful spectral structure may not be fully preserved, which likely limits the extent to which band-gap-specific information can be encoded and utilized during generation. Conversely, Bader charge, despite its compactness, may also indirectly contribute to property targeting through its encoding of local charge redistribution, which is closely related to bonding and electronic structure. Further work will therefore be needed to identify which local electronic descriptors are most suitable for joint generation and how they should be represented to retain their physically relevant information most effectively. The present results nonetheless suggest that the key criterion for selecting generative variables is not their dimensionality but their physical meaningfulness—a principle demonstrated directly by the dummy-variable control experiment. In this context, advances in machine-learning-based electronic-structure prediction \cite{deng2023chgnet} may offer useful guidance for selecting informative and learnable descriptors.

A practical implication of the proposed framework is that, because the model jointly generates Bader charges and atomic DOS alongside crystal structures, these quantities are immediately available for analysis without requiring additional DFT calculations. This offers a potential route to rapid post-generation screening and characterization of candidate materials. However, the generation accuracy of these electronic quantities tends to be higher for structures with greater thermodynamic stability and physical plausibility, as evidenced by the lower MAE observed for the SUN subset compared with the full set of generated structures. Accordingly, when using the generated Bader charges or atomic DOS for downstream analysis, it is advisable to restrict attention to structures that have been pre-screened for stability and validity using tools such as machine-learning potentials \cite{chen2022m3gnet,batatia2022mace,yang2024mattersim,park2024sevennet,deng2023chgnet} and SMACT composition-screening rules \cite{negishi2025xtalmet,davies2019smact}, so that only physically plausible candidates are analyzed. Furthermore, even within this screened subset, the accuracy of atomic DOS generation varies substantially with elemental species. As discussed above, transition-metal elements tend to yield comparatively accurate DOS profiles, whereas light nonmetallic elements exhibit larger variability. These element-dependent limitations should be taken into account when interpreting generated atomic DOS profiles for structures containing such species.

Overall, the present results suggest that jointly generating local electronic descriptors improves structure-only crystal generation by providing additional physically meaningful guidance. Its main benefit is not simply improved prediction of auxiliary electronic quantities, but a more favorable generative process that simultaneously serves two complementary roles: broadening the explored materials space through increased structural diversity, and mitigating the trade-off between property targeting and structural quality by guiding the structural trajectory toward electronically plausible configurations during joint denoising. Crucially, the dummy-variable control experiment demonstrates that these benefits require the additional variables to carry genuine physical meaning, and that expanding the generative space with uninformative variables provides no such advantage. This principle suggests that the careful selection of physically meaningful generative variables is a promising direction for further improving inverse crystal design. At the same time, the best choice and representation of electronic descriptors remain important open questions for future work.

\section{Methods}

\subsection{Dataset}

The dataset used in this study was constructed from the MP-20 dataset \cite{jain2013materialsproject}, selecting only those structures for which both the electron density and DOS were available. The resulting dataset was divided into training, validation, and test sets containing 19,571, 4,910, and 6,112 structures, respectively.

To incorporate local electronic information into the generative framework, additional site-resolved descriptors were assigned to each structure. First, Bader charges were computed for all atomic sites from the corresponding electron density using the grid-based Bader analysis algorithm \cite{bader1990atoms,henkelman2006fast,sanville2007improved,tang2009grid,yu2011accurate} and added to the dataset as local scalar quantities. For the atomic DOS, the DOS data provided for each site were directly used as the starting point for further processing.

In the DOS preprocessing, the spin channels were summed to obtain the total atomic DOS, and the energy origin was set at the Fermi energy, defined as 0~eV. The DOS was then resampled on a uniform grid spanning from $-15$ to $5$~eV with 512 points, followed by Gaussian smearing to obtain a smooth spectral representation. In addition, to suppress the contribution of semicore states as much as possible, the following truncation procedure was applied: starting from 0~eV and moving toward lower energy, if the DOS first became zero at an energy below $-5$~eV, only the spectral region above that energy was retained and the DOS at lower energies was set to zero. This treatment was introduced to remove deeply bound semicore features that are not expected to play a major role in the target electronic structure representation.

The processed atomic DOS was subsequently compressed into a 30-dimensional latent vector using an autoencoder. These latent representations were stored in the dataset for each atomic site in each structure and used as local electronic descriptors in the subsequent model development.

\subsection{Model}

We formulate generation as a joint diffusion problem over the crystal-structure variables and a site-resolved local electronic descriptor. Let a crystal be represented by $\mathcal{M}=(\bm{A},\bm{X},\bm{L})$, where $\bm{A}$ denotes atom types, $\bm{X}$ fractional atomic coordinates, and $\bm{L}$ the lattice. In addition, let local electronic descriptors be defined by $\mathcal{Z}=(\bm{Z}^0,\bm{Z}^1,\ldots)$ ($\bm{Z}^k \in \mathbb{R}^{n \times l_k}$) for the $n$ atomic sites. In the Bader-charge model, $\bm{Z}^{\mathrm{BC}} \in \mathbb{R}^{n}$ is the vector of Bader charge scalar values over atomic sites, whereas in the atomic-DOS model, $\bm{Z}^{\mathrm{ADOS}} \in \mathbb{R}^{n \times 30}$ is the matrix of latent vectors of the autoencoder-compressed atomic DOS for each site. The goal of the model is therefore to learn the joint distribution $p(\mathcal{M},\mathcal{Z})$.

Following the joint-diffusion formulation, we define a forward Markov chain $(\mathcal{M}_0,\mathcal{Z}_0) \rightarrow \cdots \rightarrow (\mathcal{M}_T,\mathcal{Z}_T)$ with the transition kernel
\begin{align}
&q(\bm{A}_{t+1},\bm{X}_{t+1},\bm{L}_{t+1},\bm{Z}_{t+1} \mid \bm{A}_t,\bm{X}_t,\bm{L}_t,\bm{Z}_t) \notag\\
&= q(\bm{A}_{t+1}\mid \bm{A}_t)\,q(\bm{X}_{t+1}\mid \bm{X}_t)\,q(\bm{L}_{t+1}\mid \bm{L}_t)\,q(\bm{Z}_{t+1}\mid \bm{Z}_t).
\end{align}
Thus, the atom types, coordinates, lattice, and local electronic descriptor are independently noised in the forward process, while their dependencies are learned jointly in the reverse process through a shared denoising network. For the structural variables, we directly inherit the MatterGen formulation: atom types are diffused by a D3PM process \cite{austin2021structured}, fractional coordinates by wrapped-normal variance-exploding diffusion, and the lattice by the custom variance-preserving diffusion \cite{zeni2025mattergen}.

For the local electronic descriptor $\bm{Z}$, we use a variance-preserving diffusion process for both the Bader-charge and atomic-DOS models. Specifically, with a noise schedule $\{\beta_t\}_{t=1}^T$ and $\bar{\alpha}_t=\prod_{i=1}^t(1-\beta_i)$, the forward process is defined as
\begin{align}
q(\bm{Z}_t \mid \bm{Z}_{t-1}) &= \mathcal{N}\!\left(\sqrt{1-\beta_t}\,\bm{Z}_{t-1}, \beta_t \bm{I}\right), \\
q(\bm{Z}_t \mid \bm{Z}_0) &= \mathcal{N}\!\left(\sqrt{\bar{\alpha}_t}\,\bm{Z}_0, (1-\bar{\alpha}_t)\bm{I}\right),
\end{align}
The reverse process is modeled as $p_{\theta}(\mathcal{M}_{t-1},\mathcal{Z}_{t-1}\mid \mathcal{M}_t,\mathcal{Z}_t,t)$, where the denoising network predicts the noise of the structural and local electronic descriptors jointly.

As the structural backbone, we adopt the original MatterGen denoising architecture based on a GemNet-T graph neural network \cite{zeni2025mattergen,gasteiger2021gemnet}. Denoting by $\bm{H}_t=(\bm{h}_{1,t},\ldots,\bm{h}_{n,t}) \in \mathbb{R}^{n \times d}$ the node features produced by the shared equivariant backbone at diffusion step $t$, a fully connected classification head maps each $\bm{h}_{i,t}$ to the categorical distribution over atom types at site $i$, while an additional node-wise fully connected regression head maps the same $\bm{h}_{i,t}$ to the predicted noise vector $\hat{\bm{\varepsilon}}^{\bm{Z}}_{\bm{\theta},i}(\mathcal{M}_t,\mathcal{Z}_t,t) \in \mathbb{R}^{l}$ for the local electronic descriptor of the corresponding atom. In this way, the structural and local electronic variables are coupled through a common message-passing representation throughout denoising.

The overall training objective is written as
\begin{align}
\mathcal{L}
= \rho_{\mathrm{coord}}\mathcal{L}_{\mathrm{coord}}
+ \rho_{\mathrm{cell}}\mathcal{L}_{\mathrm{cell}}
+ \rho_{\mathrm{types}}\mathcal{L}_{\mathrm{types}}
+ \rho_{\mathrm{local}}\mathcal{L}_{\mathrm{local}},
\end{align}
where $\mathcal{L}_{\mathrm{coord}}$, $\mathcal{L}_{\mathrm{cell}}$, and $\mathcal{L}_{\mathrm{types}}$ are the original MatterGen losses for coordinates, lattice, and atom types, respectively \cite{zeni2025mattergen}. The term $\mathcal{L}_{\mathrm{local}}$ is a weighted sum of noise-prediction losses,
\begin{align}
\mathcal{L}_{\mathrm{local}}
= \sum_{k} w_k \mathbb{E}_{\bm{Z}_0^k,\bm{\varepsilon}_{\bm{Z}^k}}
\left[
\left\|
\hat{\bm{\varepsilon}}_{\bm{Z}^k,\bm{\theta}}(\mathcal{M}_t,\mathcal{Z}_t,t) - \bm{\varepsilon}_{\bm{Z}^k}
\right\|_2^2
\right],
\end{align}
which, under the noise-prediction parameterization $\hat{\bm{\varepsilon}}_{\bm{Z}^k,\bm{\theta}} \approx \bm{\varepsilon}_{\bm{Z}^k} = \nabla_{\bm{Z}_t^k}\log q(\bm{Z}_t^k\mid\bm{Z}_0^k)$, is equivalent to minimizing a weighted sum of score-matching objectives \cite{song2019generative} across descriptor types.

In addition to this generic formulation, we introduced experiment-specific constraints during reverse sampling. Let $\tau_t = 1 - t/T$ denote the normalized reverse time, which increases from 0 to 1 during denoising process. For the Bader-charge model, charge neutrality was progressively enforced after each reverse update by
\begin{align}
\tilde{Z}_{i,t-1}^{\mathrm{BC}}
= Z_{i,t-1}^{\mathrm{BC}}
- \tau_t \frac{1}{n} \sum_{j=1}^{n} Z_{j,t-1}^{\mathrm{BC}},
\end{align}
where $Z_{i,t-1}^{\mathrm{BC}}$ denotes the Bader charge at site $i$. This correction gradually subtracts the mean excess charge from every site so that the total charge of the generated structure approaches zero, while leaving the early, highly stochastic stage of sampling relatively unconstrained. For the atomic-DOS model, each latent component is required to remain in the range $[-1,1]$, consistent with the autoencoder representation introduced in the Dataset subsection. We therefore applied the element-wise update
\begin{align}
\tilde{Z}_{ij,t-1}^{\mathrm{ADOS}}
= (1-\tau_t) Z_{ij,t-1}^{\mathrm{ADOS}}
+ \tau_t \operatorname{clip}
\!\left(
  Z_{ij,t-1}^{\mathrm{ADOS}}, -1, 1
\right),
\end{align}
so that the range restriction is weak at large noise levels and becomes strongest near the end of denoising.

\subsection{Evaluation of the V, S, U, and N Criteria}

We evaluated the generated structures using the validity (V), stability (S), uniqueness (U), and novelty (N) criteria. To maintain consistency with prior crystal-generation benchmarks, the U and N criteria were determined using the disordered structure-matching scheme provided in the MatterGen evaluation pipeline, which wraps \texttt{pymatgen}'s \texttt{StructureMatcher} and enables structure matching while accounting for disorder \cite{zeni2025mattergen,ong2013pymatgen}.

For the uniqueness criterion, generated structures were matched against one another using the disordered structure-matching scheme in MatterGen. Structures belonging to the same matched set were treated as duplicates, and only one representative from each set was counted as unique \cite{zeni2025mattergen}.

For the novelty criterion, each generated structure was compared with the public Alex--MP reference dataset released by MatterGen for novelty and stability evaluation, which contains 845,997 structures derived from MP-20 and Alexandria \cite{zeni2025mattergen,jain2013materialsproject,schmidt2023alexandria}. Structures with no match in this reference set under the same disordered matching criterion were counted as novel.

For the stability criterion, the total energies of the generated structures were obtained after structural relaxation by VASP calculations \cite{kresse1996efficient,kresse1996efficiency} using the Materials Project input parameter set (MPRelaxSet) implemented in pymatgen \cite{ong2013pymatgen} which employs the PBE exchange--correlation functional \cite{perdew1996generalized} with Hubbard +U corrections (GGA+U) for elements for which the Materials Project applies such corrections,  and the projector augmented wave (PAW) method \cite{kresse1999ultrasoft}. Using these relaxed energies, the energy above the convex hull was computed with respect to phase diagrams constructed from Materials Project data \cite{jain2013materialsproject}. A structure was counted as stable when its energy above hull was less than or equal to 0.1 eV/atom.

For the validity criterion, we considered only composition-based physical validity in the present study. Specifically, validity was evaluated using the \texttt{validity=["smact"]} option in \texttt{xtalmet}, which applies the SMACT composition-screening rules \cite{negishi2025xtalmet,davies2019smact}. Under this criterion, a composition is regarded as valid when at least one combination of oxidation states satisfies overall charge neutrality and the composition is chemically plausible in terms of the SMACT electronegativity-based heuristic \cite{davies2019smact}.

\subsection{Evaluation of Continuous Novelty Metrics}

To complement the binary novelty criterion described above, we additionally evaluated novelty using the continuous metrics proposed by Negishi \textit{et al.} \cite{negishi2025continuousnovelty,negishi2025xtalmet}. Specifically, we considered both a compositional metric and a structural metric. For the compositional metric, we used the Euclidean distance between Magpie fingerprints, which provides a continuous measure of similarity in composition space \cite{negishi2025continuousnovelty,ward2016magpie}. For the structural metric, we used the distance between average minimum distance (AMD) vectors, which provides a continuous measure of similarity in structure space \cite{negishi2025continuousnovelty,widdowson2022amd}.

In this analysis, the reference set for novelty evaluation was the MP-20 dataset. For each generated structure, we identified the nearest structure in the MP-20 reference set under each metric and used the corresponding nearest-neighbor distance as a fine-grained measure of novelty. In this way, structures that were all classified as novel by the binary criterion could be further distinguished according to how far they were from their most similar reference structures in composition space and in structure space \cite{negishi2025continuousnovelty}. This continuous analysis therefore complements the 0/1 novelty judgment by quantifying the degree of novelty relative to known materials.

\bibliographystyle{unsrt}
\bibliography{references}

\section*{Data and code availability}
The code developed for this study will be made publicly available upon publication 

\section*{Acknowledgements}
This study was supported by Japan Science and Technology Agency (JST) (Nos. JPMJAX24DB and JPMJBS2418), the Ministry of Education, Culture, Sports, Science and Technology (MEXT) (Nos. 24H00042 and 26K01205), and New Energy and Industrial Technology Development Organization (NEDO), Japan. Some computations were carried out using the computer resource offered by Research Institute for Information Technology, Kyushu University.

\clearpage
\onecolumn
\appendix

\begin{center}
{\large\bfseries
Supplementary Information for:\\[0.5em]
Inverse Materials Design via Joint Generation of Crystal Structures and Local Electronic Descriptors
}

\vspace{0.5em}

\normalsize
Ibuki Okuda$^{1}$, Izumi Takahara$^{1}$, Teruyasu Mizoguchi$^{1,2}$\\[0.5em]

\small
$^{1}$Department of Materials Engineering, The University of Tokyo\\
$^{2}$Institute of Industrial Science, The University of Tokyo
\end{center}

\section{Convolutional Autoencoder for the Compression of Atomic DOS Profile}
\subsection{Model Architecture}

\begin{figure}[H]
  \centering
  \begin{tikzpicture}[
    node distance=3.8mm,
    >=Latex,
    block/.style={draw, rounded corners, align=center, minimum height=8mm, inner sep=3pt, font=\scriptsize, text width=0.82\columnwidth},
    smallblock/.style={draw, rounded corners, align=center, minimum height=7mm, inner sep=3pt, font=\scriptsize, text width=0.72\columnwidth},
    latent/.style={draw, rounded corners, align=center, minimum height=8mm, inner sep=3pt, font=\scriptsize, text width=0.64\columnwidth},
    groupbox/.style={draw, rounded corners, inner sep=6pt}
  ]
    \node[smallblock] (input) {Input atomic DOS\\$(1,512)$};
    \node[smallblock, below=of input] (log1p) {$\log(1+x)$ transform};
    \node[block, below=of log1p] (encconv) {Encoder: 9 Conv1d blocks with GN + LReLU\\
    channels $1 \rightarrow 2 \rightarrow 4 \rightarrow 8 \rightarrow 16 \rightarrow 32 \rightarrow 64 \rightarrow 128 \rightarrow 256 \rightarrow 512$\\
    kernels $3,3,5,5,9,9,11,11,13$; strides $2,2,1,1,1,1,1,1,1$\\
    length $512 \rightarrow 256 \rightarrow 128 \rightarrow 128$};
    \node[block, below=of encconv] (prefc) {$1\times1$ Conv ($512 \rightarrow 64$) + GN + LReLU\\AdaptiveAvgPool$(32)$\\output: $64 \times 32$};
    \node[smallblock, below=of prefc] (flatten) {Flatten: $64 \times 32 = 2048$};
    \node[latent, below=of flatten] (latent) {Linear$(2048 \rightarrow 30)$ + LayerNorm\\learnable scale (\texttt{Softplus}) + \texttt{Tanh}\\$z \in (-1,1)^{30}$};
    \node[smallblock, below=5mm of latent] (fcreshape) {Linear$(30 \rightarrow 512 \times 128)$ + reshape};
    \node[block, below=of fcreshape] (decconv) {Decoder: 7 Conv1d blocks + 2 ConvTranspose1d blocks\\
    channels $512 \rightarrow 256 \rightarrow 128 \rightarrow 64 \rightarrow 32 \rightarrow 16 \rightarrow 8 \rightarrow 4 \rightarrow 2 \rightarrow 1$\\
    kernels $13,11,11,9,9,5,5,3,3$; strides $1,1,1,1,1,1,1,2,2$\\
    length $128 \rightarrow 128 \rightarrow 256 \rightarrow 512$};
    \node[smallblock, below=of decconv] (outact) {\texttt{Softplus} + $\exp(y)-1$};
    \node[smallblock, below=of outact] (output) {Reconstructed atomic DOS\\$(1,512)$};

    \draw[->] (input) -- (log1p);
    \draw[->] (log1p) -- (encconv);
    \draw[->] (encconv) -- (prefc);
    \draw[->] (prefc) -- (flatten);
    \draw[->] (flatten) -- (latent);
    \draw[->] (latent) -- (fcreshape);
    \draw[->] (fcreshape) -- (decconv);
    \draw[->] (decconv) -- (outact);
    \draw[->] (outact) -- (output);

    \node[groupbox, fit=(input)(log1p)(encconv)(prefc)(flatten)(latent), label={[font=\scriptsize]above:Encoder}] {};
    \node[groupbox, fit=(fcreshape)(decconv)(outact)(output), label={[font=\scriptsize]below:Decoder}] {};
  \end{tikzpicture}
  \caption{Compact schematic of the convolutional autoencoder used for atomic-DOS compression. The model maps a 512-point one-dimensional spectrum to a 30-dimensional latent vector and reconstructs the spectrum through a mirrored decoder. GN and LReLU denote GroupNorm and LeakyReLU, respectively.}
  \label{fig:autoencoder_architecture}
\end{figure}
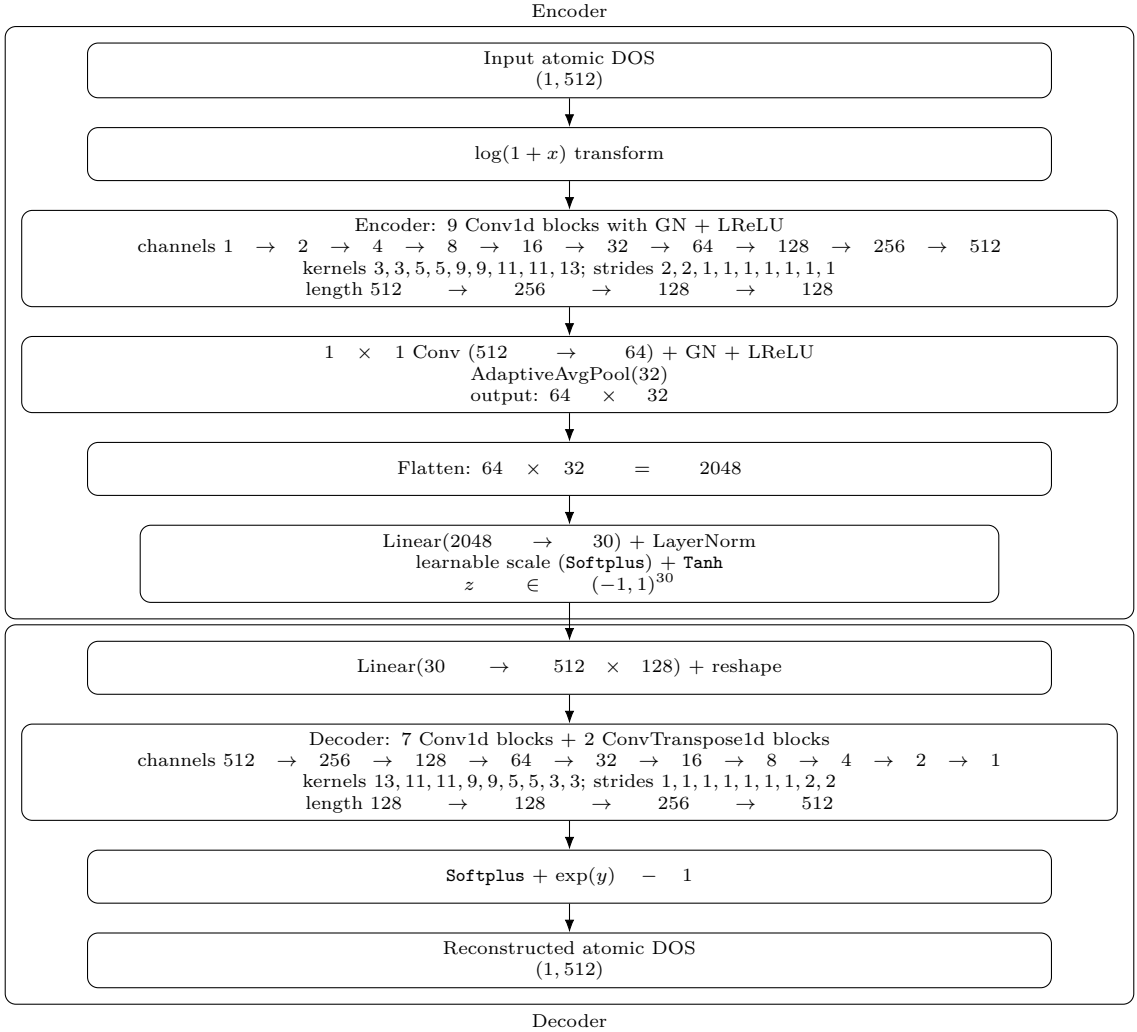

To obtain a compact site-resolved electronic descriptor for the atomic DOS, we used the one-dimensional convolutional autoencoder. The model takes a preprocessed atomic DOS spectrum sampled on 512 energy points as a one-channel input signal of shape $(1,512)$, encodes it into a 30-dimensional latent vector $z \in (-1,1)^{30}$, and decodes it back to a reconstructed spectrum of the same length. This bounded latent representation is consistent with the range-constrained diffusion formulation described in the main text. The encoder first applies a $\log(1+x)$ transform to compress the dynamic range of the spectral intensity, whereas the decoder restores the physical scale by applying \texttt{Softplus} followed by the inverse transformation $\exp(y)-1$ at the output layer.

The encoder consists of nine one-dimensional convolutional blocks (\texttt{Conv1d} + GroupNorm + LeakyReLU). The first two blocks downsample the spectral length from 512 to 128, while the remaining blocks keep the length fixed and expand the channel dimension to build a high-capacity representation. The resulting feature map is compressed by a $1\times1$ convolution and adaptive average pooling, flattened, and projected to a 30-dimensional latent vector. A LayerNorm, a learnable scalar rescaling (\texttt{Softplus}), and a final \texttt{Tanh} constrain the latent components to $(-1,1)$.

The decoder inverts this structure: the latent vector is linearly expanded and reshaped into a feature map, passed through convolutional blocks with LeakyReLU activations at fixed length, and then upsampled by two transposed convolutions to recover the original 512-point resolution. The output is passed through \texttt{Softplus} and the inverse \texttt{log1p} transform, $\exp(y)-1$, to reconstruct the DOS on its original scale. This architecture provides a compact but information-preserving latent representation suitable for downstream diffusion modeling.

\subsection{Loss Function}

The autoencoder was trained using a composite reconstruction loss designed to balance fidelity in the original intensity scale, robustness in the logarithmically compressed space, and preservation of spectral shape. Let $x \in \mathbb{R}^{512}_{\ge 0}$ denote the target atomic DOS and $\hat{x}$ its reconstruction. In the actual training code, the objective is implemented as
\begin{align}
\mathcal{L}
= \lambda_{\mathrm{abs}} \mathcal{L}_{\mathrm{abs}}
+ \mathcal{L}_{\mathrm{log}}
+ \lambda_{\mathrm{rel}} \mathcal{L}_{\mathrm{rel}}
+ \lambda_{\mathrm{grad}} \mathcal{L}_{\mathrm{grad}},
\end{align}
with $\lambda_{\mathrm{abs}} = 10^{-3}$, $\lambda_{\mathrm{rel}} = 0.5$, and $\lambda_{\mathrm{grad}} = 0.1$. Each term uses the Smooth L1 loss with parameter $\beta = 1.0$, which provides a robust interpolation between the $\ell_1$ and $\ell_2$ regimes and reduces sensitivity to occasional large reconstruction errors.

The first term measures the reconstruction error directly in the original DOS scale,
\begin{align}
\mathcal{L}_{\mathrm{abs}} = \mathrm{SmoothL1}(\hat{x}, x).
\end{align}
Because the DOS may vary over a wide dynamic range, the main reconstruction term is instead evaluated in the log-transformed space,
\begin{align}
\mathcal{L}_{\mathrm{log}}
= \mathrm{SmoothL1}\!\left(\log(1+\hat{x}), \log(1+x)\right).
\end{align}
This term emphasizes agreement over both high- and low-intensity regions after dynamic-range compression and is therefore better suited to spectral data than a purely linear-scale loss.

To further account for differences in overall magnitude across samples, the training objective includes a relative reconstruction term. Defining
\begin{align}
m(x) = \max_j x_j,
\end{align}
with a small lower bound for numerical stability in the implementation, the normalized spectra are compared as
\begin{align}
\mathcal{L}_{\mathrm{rel}}
= \mathrm{SmoothL1}\!\left(
\log\!\left(1+\frac{\hat{x}}{m(x)}\right),
\log\!\left(1+\frac{x}{m(x)}\right)
\right).
\end{align}
This term suppresses sample-to-sample amplitude variation and encourages the model to reproduce relative spectral profiles even when the absolute DOS scale differs.

In addition, a multi-scale gradient-matching term is introduced to preserve local peak shapes and edge-like spectral features. For a smoothing scale $s$, the implementation first applies average pooling with kernel size $s$ and then evaluates a centered finite difference. Denoting this discrete gradient operator by $g_s(\cdot)$, the gradient loss is written as
\begin{align}
\mathcal{L}_{\mathrm{grad}}
&= \frac{1}{|\mathcal{S}|}
\sum_{s \in \mathcal{S}}
\,\mathrm{SmoothL1}\!\bigg(
g_s\!\left(\log\!\left(1 + \frac{\hat{x}}{m(x)}\right)\right), g_s\!\left(\log\!\left(1 + \frac{x}{m(x)}\right)\right)
\bigg),
\end{align}
where $\mathcal{S} = \{1,2,4\}$ in the present implementation. By combining gradient matching across multiple scales, the loss encourages consistency not only in overall spectral weight but also in the positions and sharpness of fine and moderately broadened features.

Taken together, the loss design reflects the requirements of atomic-DOS compression more faithfully than a single pointwise error. The direct and log-domain terms stabilize the reconstruction of amplitudes across a broad dynamic range, the relative term reduces the effect of sample-dependent overall intensity, and the gradient term improves the recovery of peak structure.

\section{Element-Resolved Generation Accuracy}

\begin{figure*}[t]
  \centering
  \includegraphics[width=\textwidth]{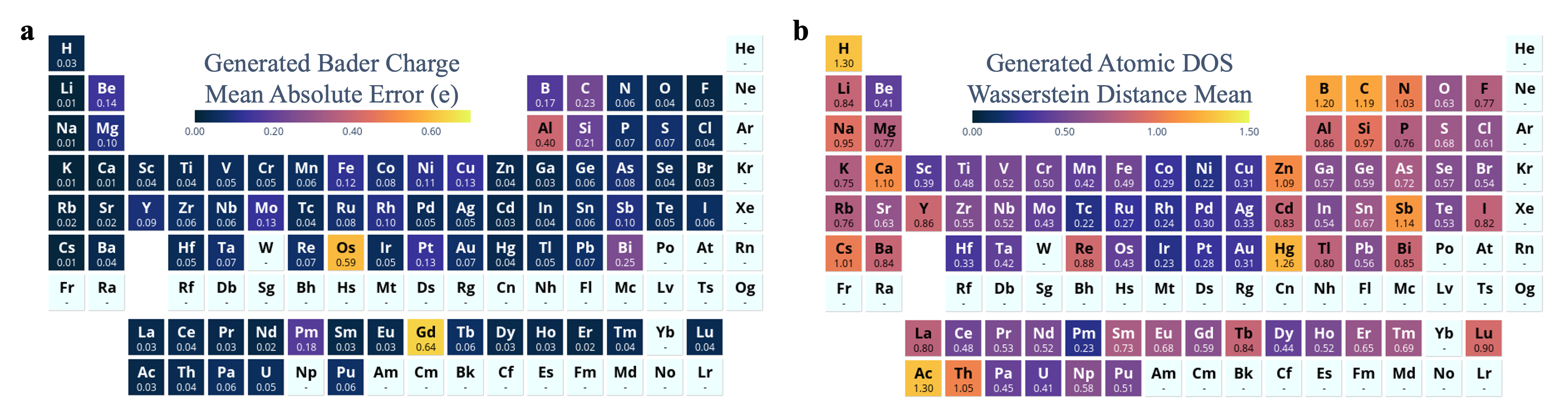}
  \caption{
    Element-resolved generation accuracy visualized on the periodic table.
    \textbf{a.}~Mean absolute error (MAE) of the generated Bader charges relative to DFT-calculated reference values, evaluated for each element over all generated structures in the \textit{ab initio} setting.
    \textbf{b.}~Mean Wasserstein distance $W_1$ between generated and DFT-calculated atomic DOS profiles, evaluated for each element under the same setting.
  }
  \label{fig:accuracy_periodic}
\end{figure*}

Figure~\ref{fig:accuracy_periodic} shows the element-resolved generation accuracy for both Bader charge and atomic DOS, visualized on the periodic table.

For Bader charge (Fig.~\ref{fig:accuracy_periodic}\textbf{a}), the MAE remains below $0.1\,e$ for the majority of elements, including alkali and most alkaline-earth metals. The largest deviations are observed for Os ($0.59\,e$) and Gd ($0.64\,e$), both of which are scarcely represented in the training set, appearing at only $315$ ($0.4\%$) and $483$ ($0.6\%$) atomic sites, respectively. Somewhat elevated values are also found for the light $p$-block elements B, C, and Si ($0.17$--$0.23\,e$).

For atomic DOS (Fig.~\ref{fig:accuracy_periodic}\textbf{b}), the open-$d$-shell transition metals across the $3d$, $4d$, and $5d$ series consistently exhibit low Wasserstein distances ($W_1 \approx 0.2$--$0.5$), in agreement with the spectral regularity of $d$-derived bands discussed in the main text. The largest $W_1$ values, in contrast, are concentrated at hydrogen ($1.30$) and the light $p$-block elements B, C, and N ($1.03$--$1.20$), which exhibit a broad range of bonding characters across compounds, from covalent $sp$-hybridized to predominantly ionic, giving rise to diverse spectral profiles.

Taken together, the two panels reveal that the learnability of local electronic descriptors depends on chemical identity in qualitatively distinct ways for the two descriptor types. For Bader charge, accuracy is high across most of the periodic table, and the few elements with elevated MAE are primarily those sparsely represented in the training data. Atomic DOS generation, in contrast, shows a systematic dependence on the chemical character of each element: elements whose local electronic structure is more sensitive to the surrounding bonding environment yield consistently larger errors, regardless of their abundance in the training set.

\balance
\end{document}